\documentclass{article}

\usepackage{natbib,epsfig,latexsym,amsmath,amssymb}

\begin{document}

\title{Inference of genetic networks from time course expression data using functional regression with lasso penalty}
\author{Heng Lian\\
Division of Mathematical Sciences,\\ School of Physical and Mathematical Sciences,\\ Nanyang Technological University, 637371,\\ Singapore}

\maketitle

\begin{abstract}

Statistical inference of genetic regulatory networks is essential for understanding temporal interactions of regulatory elements inside the cells. For inferences of large networks, identification
of network structure is typically achieved under the assumption of sparsity of the networks. However, current approaches either have difficulty extending to networks with a large number of genes due
to the computational constraints, or are difficult to interpret due to the use of module-based models. Also, in most previously proposed models, the choice of different parameters in the model is dealt with in a heuristic manner. For example, in LEARNe \citep{nam07}, which used a system of differential equations to model the dynamics, an upper bound for the degree of connectivity of the network is assumed to be known.

When the number of time points in the expression experiment is not too small, we propose to infer the parameters of the ordinary differential equations using the techniques from functional data analysis (FDA) by regarding the observed time course expression data as continuous-time curves. The derivative of the expression curve with respect to time is easily calculated without using finite difference and thus the problem of missing observations or unequally spaced time points can be dealt with in a consistent way in this model. For networks with a large number of genes, we take advantage of the sparsity of the networks by penalizing the linear coefficients with a $L_1$ norm. The model is fitted using the efficient algorithm that finds the whole regularization path of the parameters which in turn produces the whole spectrum of possible performances in terms of sensitivity and positive
predictive value (PPV). The smoothing parameters can be chosen via cross-validation which gives a reasonable trade-off between PPV and sensitivity. The algorithm is compared to the state-of-the-art program LEARNe in simulations for small to medium sized networks which shows the competitiveness of the new approach. The ability of the algorithm to infer network structure is demonstrated using the cell-cycle time course data for \textit{Saccharomyces cerevisiae}.

\end{abstract}

\section{Introduction}
The increasing amount of high-throughput time course data has
provided biologists a window to the understanding of the
biomolecular mechanism of different species. The expression of
genes in these studies are indicative of the dynamic activities
occurring inside the organism. Such regulatory activities involve
complicated temporal interactions among different gene products,
forming genetic networks indicating the causal relationships
between different elements. It is the responsibilities of the
statisticians to construct such networks  using statistical
models that uncovers such relationships. The utility of such
models would be vital for the discovery of biological processes
that are crucial for understanding of interactions of molecules involved in drug responses.

Different models have been proposed for the construction and
analysis of such networks from time course data, including
stochastic differential equations \citep{chen05} and graphical models \citep{friedman04}. Bayesian
methods can be applied which explore hidden causal relationships
between different nodes. In particular, dynamic Bayesian networks \citep{yu04} has achieved great success. For optimization of the networks, a large sample size is required for accurate estimation
of the structure and the amount of computations required increases
rapidly with the number of nodes,
thus limiting the application to networks with a small number of genes.

Due to the obvious connection of the problem to the traditional
time series analysis, multivariate autoregressive model has been
used to fit the expression data. The utility of this
model is constrained by the length of the time course data which
is typically on the order of tens of times points, while the number of genes is much larger. Classical maximum
likelihood estimator cannot be applied when the number of genes is
greater than the number of time points.

To overcome the problem, several algorithms have been introduced,
almost all of which applied some dimensionality reduction techniques.
Subset selection \citep{gardner03} proceeds by searching for a subset of nodes with a fixed size that
minimizes the least mean square error and use this subset for
model fitting and inference. This typically leads to poor
generalization performance due to overfitting because the number
of time points is small. In addition, one needs to choose the size
of the subset for searching and least square errors for subsets with different sizes are
not directly comparable to each other. This choice would depend on one's
belief about the degree of connectivity of the networks, which is
difficult to assess a priori in applications, and choosing a too small subset size
 obviously is disastrous to the performance. Alternatively,
singular value decomposition (SVD), utilizing a
parsimonious set of features underlying the data, can be used to
reduce the dimension and infer the networks \citep{bansal06}, but simulations showed
that its performance is sensitive to the number of features
selected, which is difficult to determine in practice. Based on similar principles,
the state space approach uses low dimensional latent variables to
reduce the number of parameters to be estimated, but the resulting
latent variables are difficult to interpret \citep{hirose08}. The difficulty in
interpreting the model also originates from the fact that the estimated 
networks do not have the sparsity property. Besides, it is
difficult to apply the state space model to time course data with
unequally spaced time points.

To address the previously mentioned overfitting effect of subset selection,
\citep{nam07} proposed combining multiple models with least mean square error below a certain threshold. The motivation comes from the machine
learning literature where model averaging or multiple voting is
often observed to improve generalization performance. It was shown
that this approach significantly outperforms simple subset selection and is at least
comparable and sometimes better than SVD even when the number of features in
SVD is optimally chosen in different situations, except when
the number of time points is  extremely small (six or smaller).
The resulting algorithm named LEARNe, however, shares one common
disadvantage with subset selection: an exhaustive search over all possible
subsets with a fixed size is conducted to find the good
performers, which is infeasible when the number of genes is large.
Additionally, some arbitrary threshold should be used to find the final
connectivity structure of the networks. Due to the heuristic
nature of the algorithm, no statistical theory seems to exist for
the choice of this threshold.

All the above mentioned approaches directly used expression data
at discrete time points for fitting the model, which might be
undesirable when the noise level is high. This is especially true when we
use the ordinary differential equations to model the networks which
requires estimation of the derivatives. In both LEARNe and SVD
algorithms, the derivatives are replaced by the difference of
expression level between consecutive time points, which might be a
poor estimate of the derivative.

In this article, we propose a novel algorithm, using a penalized form of functional regression, by
modelling the time course expressions levels over a certain period as
continuous curves after smoothing the expression data. The sparsity
of the networks is enforced by introducing $L_1$ penalty on the
constant coefficients of the differential equations using another
smoothing parameter. By varying the smoothing parameter, we can
trace out the whole performance curve of our algorithm, which can
be done using the modification of the least angle regression
(LARS) program \citep{efron04}. The sparsity of the network can be inferred based
on the data using cross-validation (CV) if desired. Unlike
LEARNe, the algorithm is very efficient in computation and can
deal with graphs with several hundred nodes when implemented in
R on a personal computer. Our simulation shows that the algorithm has comparable and sometimes better performance
than LEARNe, while sparsity is obtained without choosing an
arbitrary threshold for the coefficient matrix.

\section{Methods}
As a first step, we need to convert the time course expression
data to continuous curves. This problem falls into the realm of
functional data analysis (FDA) as studied extensively in the monograph
\citep{ramsey05}. We use $g_{ij}, i=1,\ldots,n, j=1,\ldots,n_i$ to denote the
expression level of gene $i$ at time points $t_j$ with $0\le
t_1<\ldots <t_{n_i}\le 1$. These expression levels have possibly been
preprocessed and log transformed which usually results in better
fit. Separately for each gene $i$, we search for the smooth
function $g_i(t)$ that minimizes the following functional,
\[\frac{1}{n_i}\sum_j(g_{ij}-g_i(t_j))^2+\lambda_1\int (g_i''(t))^2\,dt.\]
In the above, the first term enforces the closeness of the curve to
the observed expression level at the discrete time points, the
second term enforces the smoothness of the function $g_i$ with
larger smoothing parameter $\lambda_1$ resulting in a smoother
function. Note that with this approach, we don't need to assume
equally spaced time points or identical time points for different
genes.

In practice, the above optimization is perform by assuming $g_i$
has an expansion in terms of a certain basis
\[g_i(t)=\sum_{j=1}^K a_{ij}b_j(t).\]
After plugging in the above expansion, we only need to solve the $K$
dimensional parameters vector $\{a_{ij}\}$, which is an easy
convex optimization problem.

The perhaps most popular basis used in this context is the
B-spline basis with order $4$ (i.e. cubic splines). $K$ can be
chosen large enough while smoothness of the function is control by
the smoothing parameter $\lambda_1$. The automatic choice for
$\lambda_1$ can be made using statistical methods like cross-validation. In our experience, we find that the result is quite
robust to the choice of this parameter and we find it more
convenient to use a fixed parameter in all experiments.

In general, the dynamics of regulatory networks can be written
nonparametrically as
\[g'_{i}(t)=f(g_1(t),\ldots,g_n(t)), \,t\in[0,1]\]
for a network with $n$ nodes. Inference of such general
nonparametric model is difficult with limited amount of data. As a
first order approximation, same as \citep{nam07}, we model the
regulatory networks using the system of linear ordinary
differential equations
\[g'_{i}(t)=\alpha_i+\sum_{j=1}^n\beta_{ij}g_j(t), \, t\in [0,1], \,i=1,\ldots,n\]
with $\beta_{ij}$ representing the regulatory effects of gene $j$
on gene $i$. The interpretation is that gene $j$ activates gene
$i$ if $\beta_{ij}>0$ and gene $j$ depresses gene $i$ if
$\beta_{ij}<0$.

From the estimate of $g_i(t)$, the derivative can be easily
evaluated by $g'_i(t)=\sum_j a_{ij}b'_j(t)$. The network
coefficients $\alpha_i, \beta_{ij}$ can be fitted by minimizing
\[\int_0^1(g'_i(t)-\alpha_i-\sum_j\beta_{ij}g_j(t))^2dt\]
Unlike the discrete least squares, even when the original time
points is smaller than the number of genes, there usually exists a
unique minimizer of the above problem. However, overfitting still
occurs when the number of genes is large. From biological
considerations, the network is usually sparse with the evolution of
the expression level of one gene only depending on the expression
levels of a few other genes, which implies most of the interaction
coefficients $\beta_{ij}$ are actually zero. The $L_1$-norm
penalty, also commonly called lasso penalty, is well-known to
produce sparse regression coefficients \citep{tibshiranni96}. Despite its popularity, we
are unaware of its previous application in functional data
analysis.

With the lasso penalty added, we will optimize
\[\int_0^1(g'_i(t)-\alpha_i-\sum_j\beta_{ij}g_j(t))^2dt+\lambda_2|\beta_i|_1\]
where $\beta_i=\{\beta_{i1},\ldots,\beta_{in_i}\}$ is the network
coefficients and $\lambda_2$ is a smoothing parameter with larger
$\lambda_2$ producing sparser networks. By varying the smoothing
parameter, we can produce a whole spectrum of networks with
different degrees of sparsity. We approximate the integral
$\int_0^1(g'_i(t)-\alpha_i-\sum_j\beta_{ij}g_j(t))^2dt$ by the
discretized
\[\frac{1}{T} \sum_{t=1}^T (g'_i(\frac{t}{T})-\alpha_i-\sum_j\beta_{ij}g_j(\frac{t}{T}))^2\]
$T$ is chosen to be large enough to approximate the integral well,
and we find $T=20$ is sufficient for our simulations.

After discretization, the optimization problem becomes a standard linear regression
with lasso penalty, which can be solved after converting to a
quadratic programming problem \citep{tibshiranni96}. Computation with different values
of $\lambda_2$ makes this algorithm less efficient. Fortunately,
there exists an algorithm that computes the whole regularization
path for the coefficients for all values of the smoothing
parameter $\lambda_2$ which makes our approach very efficient
computationally. This algorithm is a modification of least angle
regression and takes advantage of the fact that the solution path
is piecewise linear \citep{efron04,rosset07}.

If one desires to choose the parameter $\lambda_2$ based on the
data, we can either use cross-validation within each gene which
results in a different smoothing parameter for each gene, or we
can use cross-validation on the whole dataset, which produces a
single smoothing parameter for all the genes. Since the lasso
penalty shrinks many networks coefficients to zero, we can infer
the sparsity of the networks based on the data without using
arbitrary thresholds for the coefficients as is commonly done in
previous approaches when the structure of the network is unknown.

Since we will compare our approach to LEARNe, we will briefly
describe that algorithm here. In LEARNe, one performs a least
square regression for each subset $S$ of $\{1,\ldots,n\}$ of size
$k$. That is, for each fixed gene $i$, one minimized
\[\sum_{j=1}^{n_i-1} (\Delta g_{ij}-(\alpha_i+\sum_{s\in S}\beta_{is}g_{sj})\Delta_{j})^2\]
where $\Delta g_{ij}=g_{i(j+1)}-g_{ij}$ and
$\Delta_{j}=t_{j+1}-t_j$. Thus LEARNe uses finite difference
method to approximate the derivative of the expression level with
respect to time. Each possible subset $S$ corresponds to a
different linear regression model. Instead of using one single
best model, which will overfit the data with a small number of
observations, one collects the top $\mu\%$ models with the
smallest sum of square above. This represents all the models that
can fit the data reasonably well. Each model will vote
independently on the signs of the network coefficients and the
votes are collected into a $n\times(n+1)$ matrix $\Theta$, whose
entries are integers with a large positive integer indicating a
strong activating effect and large negative integer indicating a
strong repressing effect. The final model is found with a
thresholding procedure on $\Theta$. If the coefficients are
desired, it can be calculated by least square regression with the
final model. No suggestion was provided for choosing the threshold in \citep{nam07}.

The authors of \citep{nam07} found by simulations that the result is robust to the
choice of $\mu$ and the method consistently outperformed subset
selection using a single model with the smallest least square error. It
is also better than SVD unless the number of time points is
unreasonably small. The most serious drawback of LEARNe in our
opinion is the computational burden when $n$ is large. The author
used $k=4$ in their simulation and this results in searching among
${n\choose 4}$ models. For example, when $n=100$, it contains
close to 4,000,000 possible models! Empirically, even for $n=50$
with $k=2$, we find our algorithm is much faster than LEARNe,
although this might be attributed to our poor implementation of
LEARNe.

In \citep{nam07}, no discussion is offered on the choice of $k$. This
value obviously should depend on the sparsity of the networks. In
practise, since the connectivity of the networks is 
unknown, it is difficult to choose appropriate $k$ especially
considering the computational complexity that comes with large
$k$.


%
\section{Results}
We compare the performance of our functional analytical approach with LEARNe in simulations. The
networks are generated as follows. For an even number of genes
$n=2r$ and $m$ time points $\{1/m,2/m,\ldots,1\}$, The $n\times
n$ coefficient matrix $A$ is generated as follows.
\[A_{2i-1,2i-1}=A_{2i,2i}=a_i, A_{2i-1,2i}=-A_{2i,2i-1}=b_i, \]
\[A_{i,j}=0 \mbox{ all other } i,j\]
\[a_i\stackrel{iid}{\sim} Uniform(-2,0), b_i\stackrel{iid}{\sim} Uniform(-5,5)\]
The structure of $A$ has the form
\[A=\left[\begin{array}{ccccccc}
a_1&b_1&0&0&\cdots&&\\
-b_1&a_1&0&0&\cdots&&\\
0&0&a_2&b_2&\cdots&&\\
0&0&-b_2&a_2&\cdots&&\\
\vdots&\vdots&\vdots&\vdots&\ddots&&\\
&&&&&a_r&b_r\\
&&&&&-b_r&a_r
\end{array}\right]\]
and the system of differential equations is written in matrix form
\[G'=AG\]
with $G(t)=(g_1(t),\ldots,g_n(t))^T$. Note the coefficients $\alpha_i$ do not appear in this simulation and also not used when fitting the model.

We generate the initial expression level $G(0)$ from Uniform
distribution and solve the initial value differential equation
problem using simple Euler method. The solution $G$ is evaluated
at those $m$ time points and independent normal noise with
variance $\sigma^2$ is added at each time point.

By the data generation mechanism, the evolution of one gene only
depends on the expression level of itself as well as one other
gene. The coefficient values $a_i$ are chosen to be negative so
that the solution of the differential equations is asymptotically
stable to avoid numerical problems.

We used several combinations of parameters $n,m,\sigma$ for our
simulation. For each combination, we use 50 randomly generated
time course expression matrix and calculate the average
performance over these data. In our algorithm, by vary
$\lambda_2$, we can reconstruct networks of varying degree of
sparsity. By counting the number of connections in the
reconstruction and the true network, the performance can be measured
using the positive predictive value (PPV) versus sensitivity
plots.

\begin{figure}[!tpb]
\center{(a)}
\centerline{\includegraphics[width=6cm]{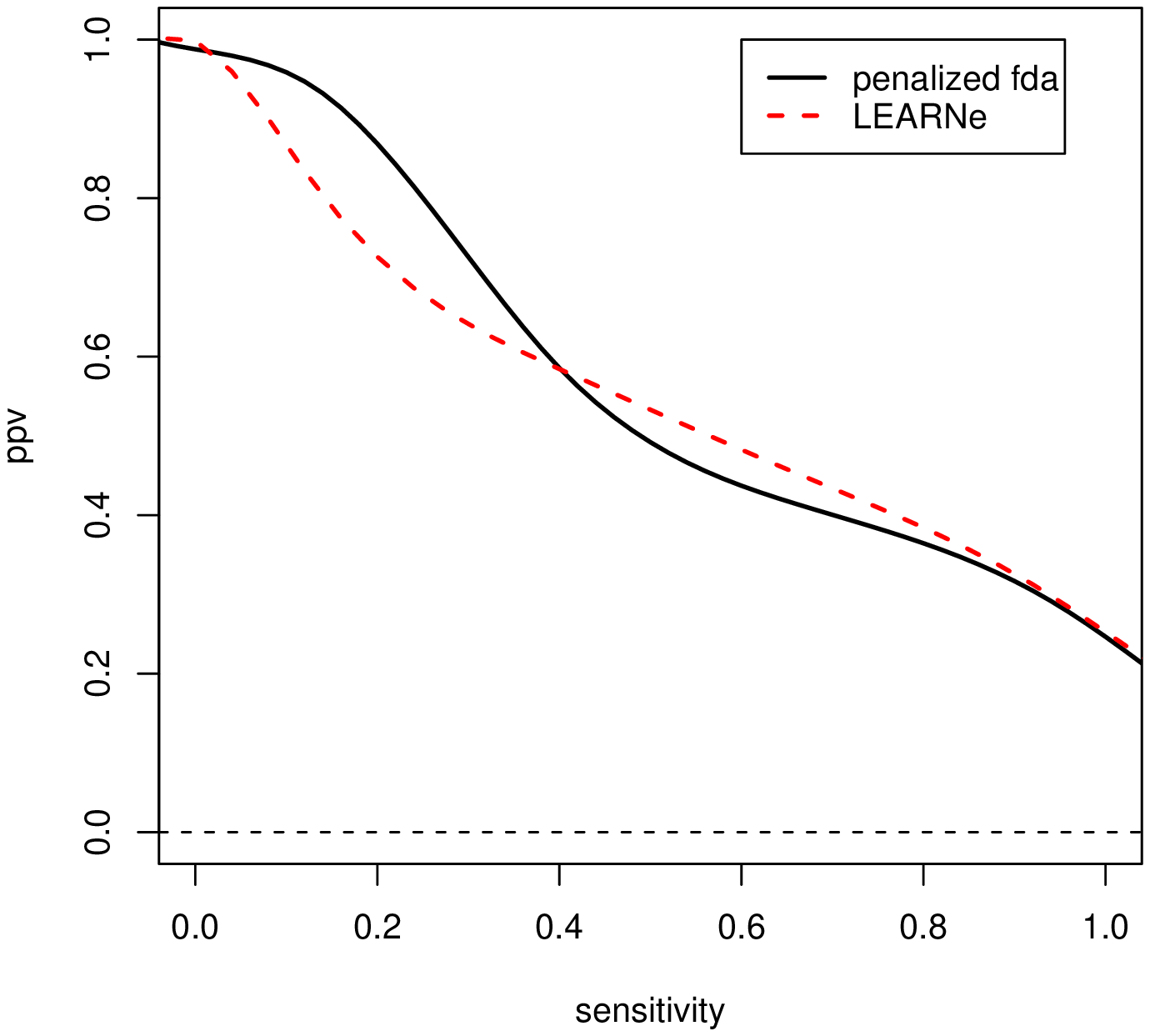}}
\center{(b)}
\centerline{\includegraphics[width=6cm]{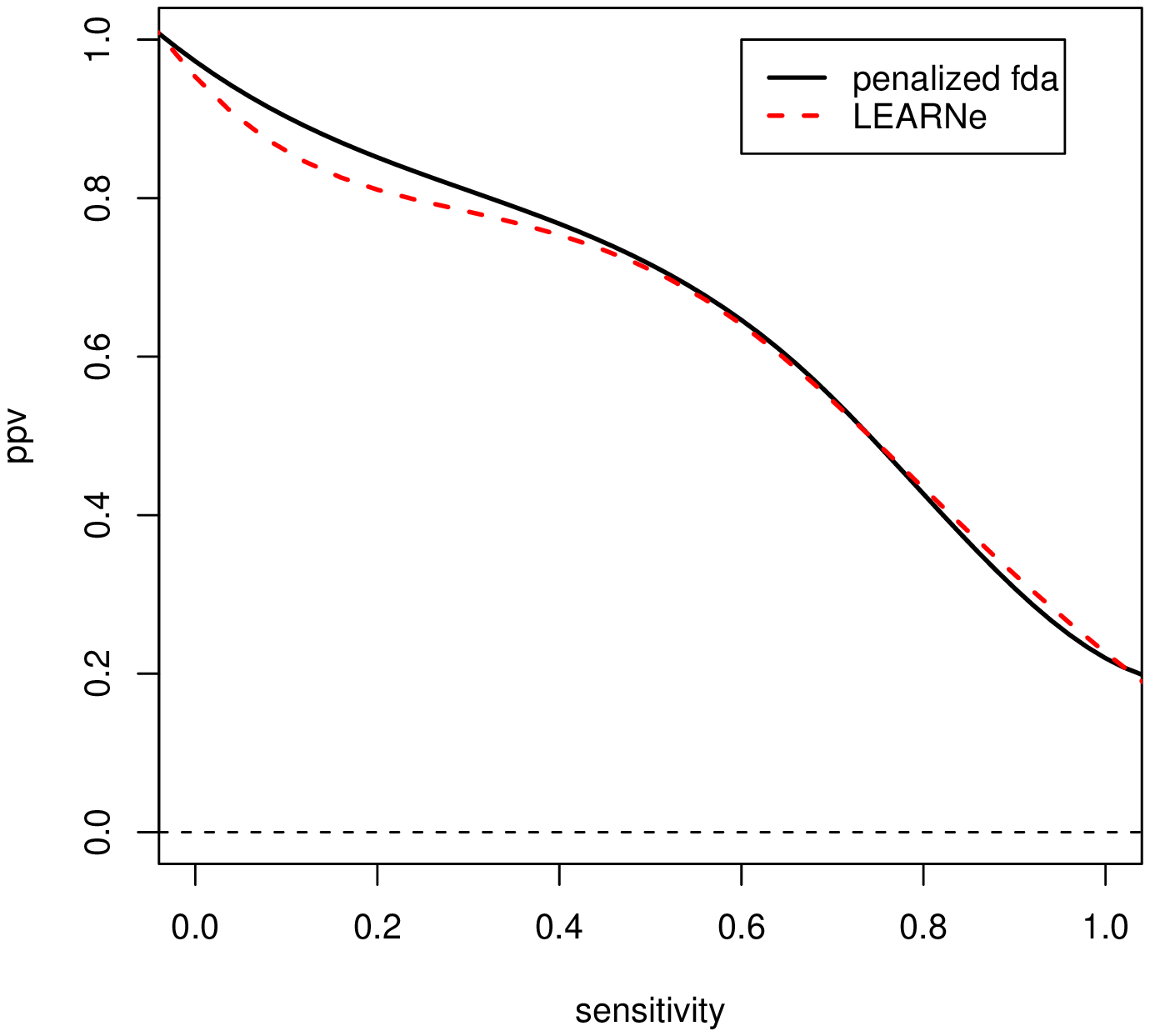}}
\center{(c)}
\centerline{\includegraphics[width=6cm]{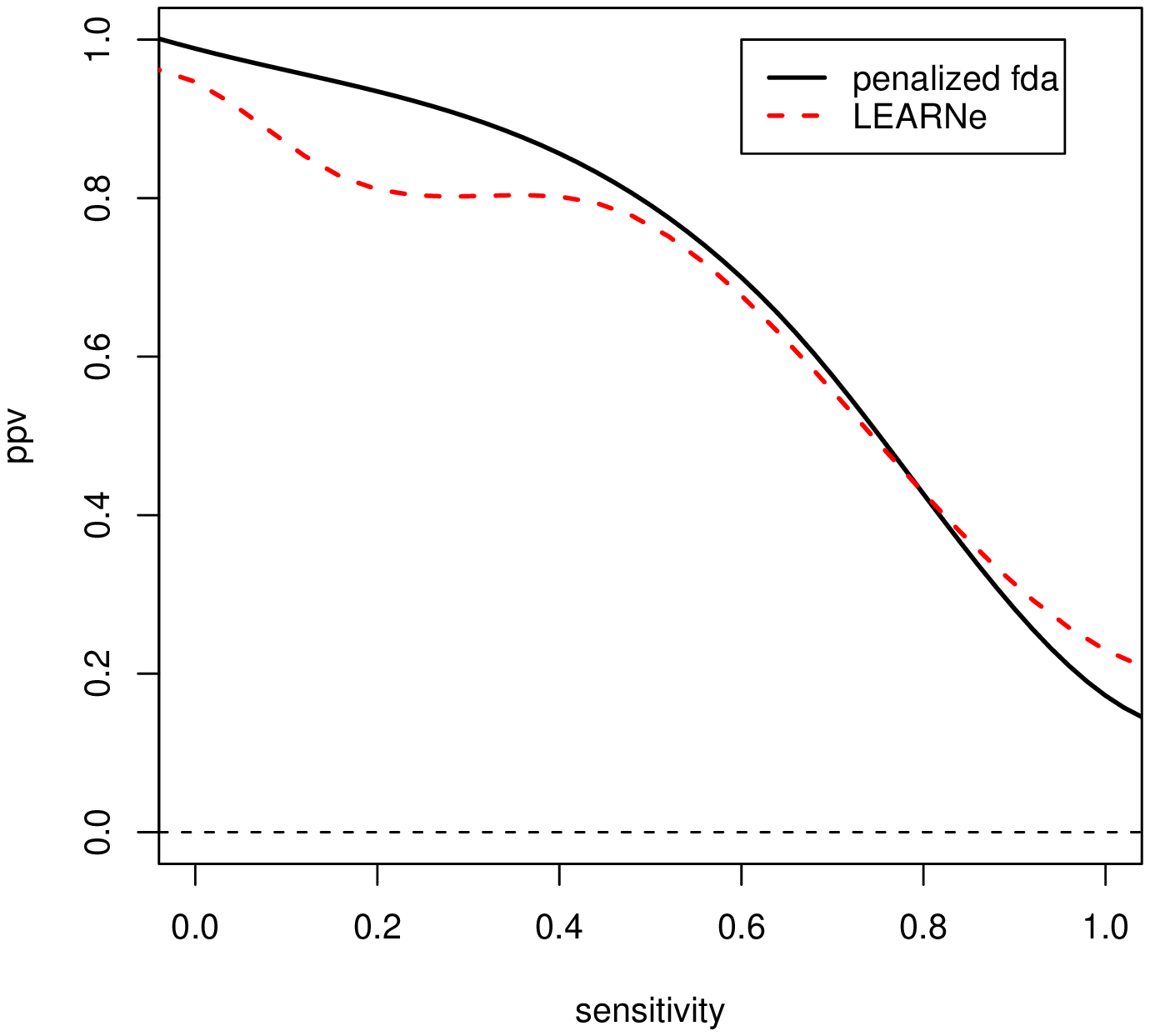}}
\caption{PPV vs. sensitivity plots for simulated networks with 10
nodes and (a) 10 (b) 30 (c) 50 time points. The standard deviation
of the noise is $\sigma$=0.1 }\label{fig:lownoise}
\end{figure}

\begin{figure}[!tpb]
\center{(a)}
\centerline{\includegraphics[width=6cm]{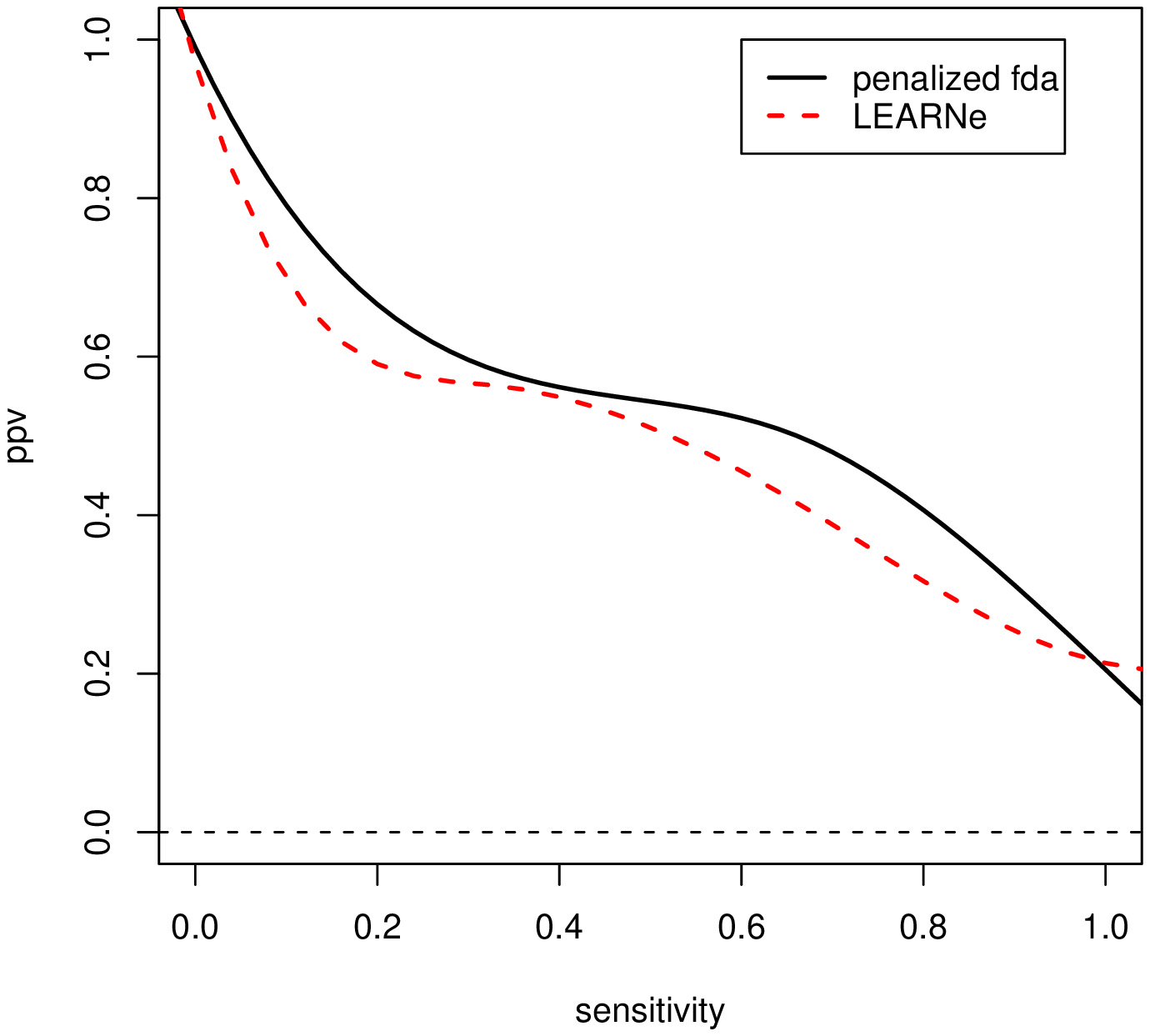}}
\center{(b)}
\centerline{\includegraphics[width=6cm]{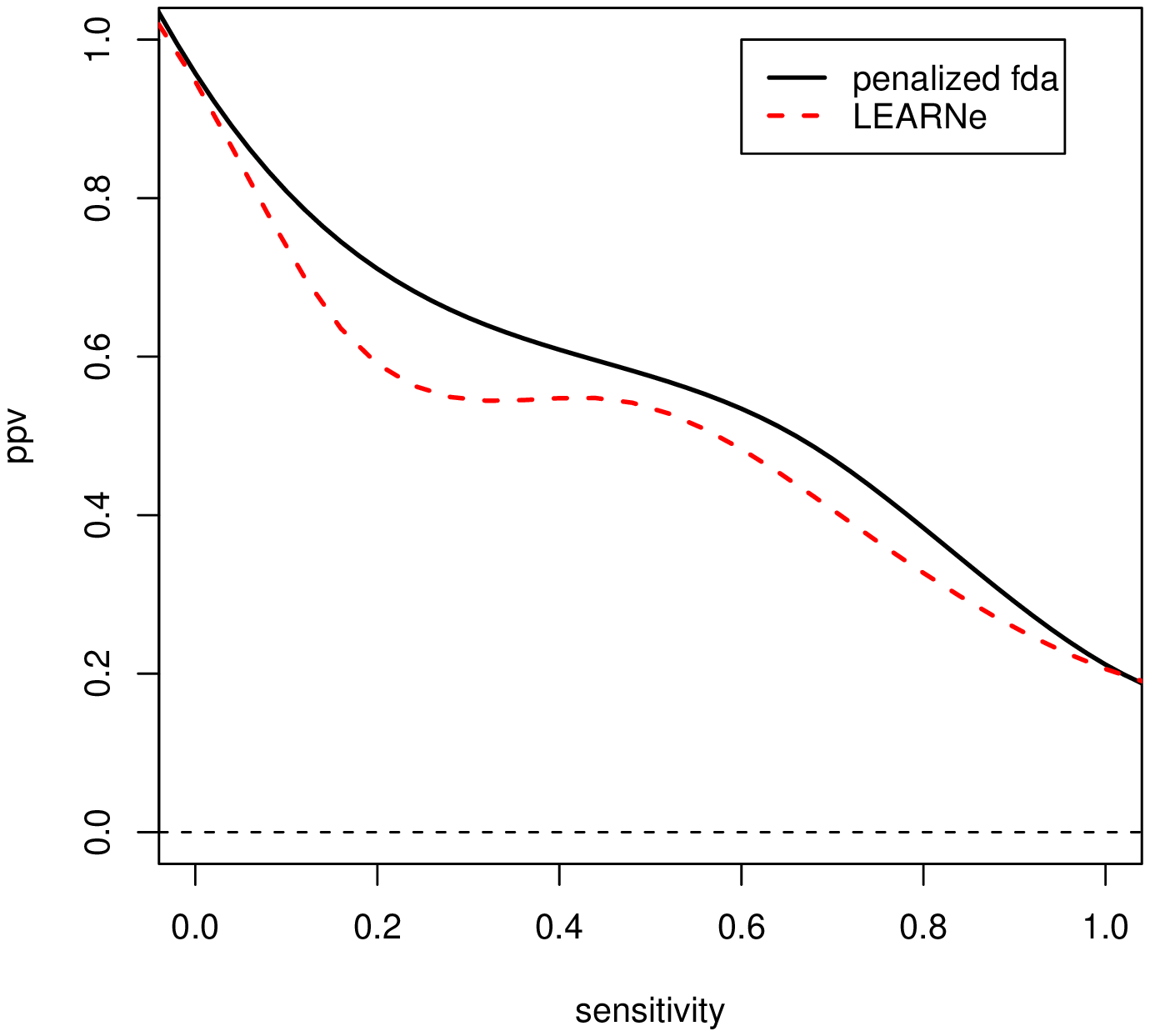}}
\center{(c)}
\centerline{\includegraphics[width=6cm]{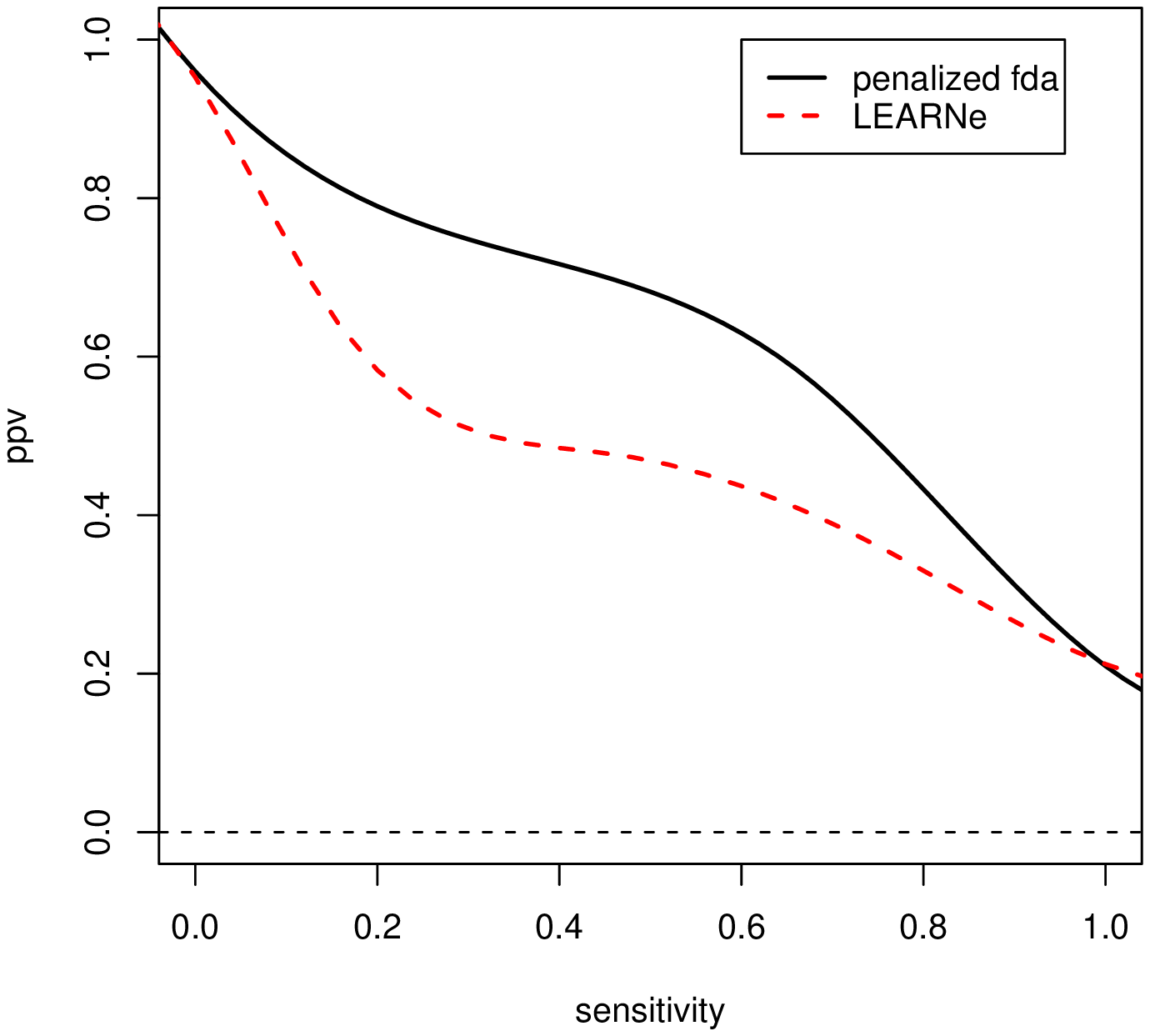}}
\caption{PPV vs. sensitivity plots for simulated networks with 10
nodes and (a) 10 (b) 30 (c) 50 time points. The standard deviation
of the noise is $\sigma$=0.3.}\label{fig:highnoise}
\end{figure}

The simulation results are shown in Figure \ref{fig:lownoise} and Figure
\ref{fig:highnoise}. These curves are produced by averaging over 50 randomly
generated data sets with smoothing for visualization. That is,
each curve is the  result of using nonparametric smoothing
over 50 PPV vs. sensitivity curves after applying a particular
algorithm, either penalized functional approach or LEARNe. From
those figures, it can be seen that for networks with 10 nodes,
when the number of time points is small and with small noise
level  the performance of our algorithm is similar to LEARNe. But for simulation with a larger number of time
points and larger noise level, the performance of our algorithm
 becomes better than LEARNe. This is possibly due to
the fact that the finite different approximation to derivatives
used in LEARNe comes into trouble in these situations. We also performed
simulations with $n=30$ and $n=50$ and observed similar effects.
More importantly, we are able to run our algorithm on networks
with $n=500$ (taking about 20 minutes) while it is impossible to
run LEARNe with $n$ bigger than $100$ in our implementation.

\begin{figure}[!tpb]
\centerline{\includegraphics[width=7cm]{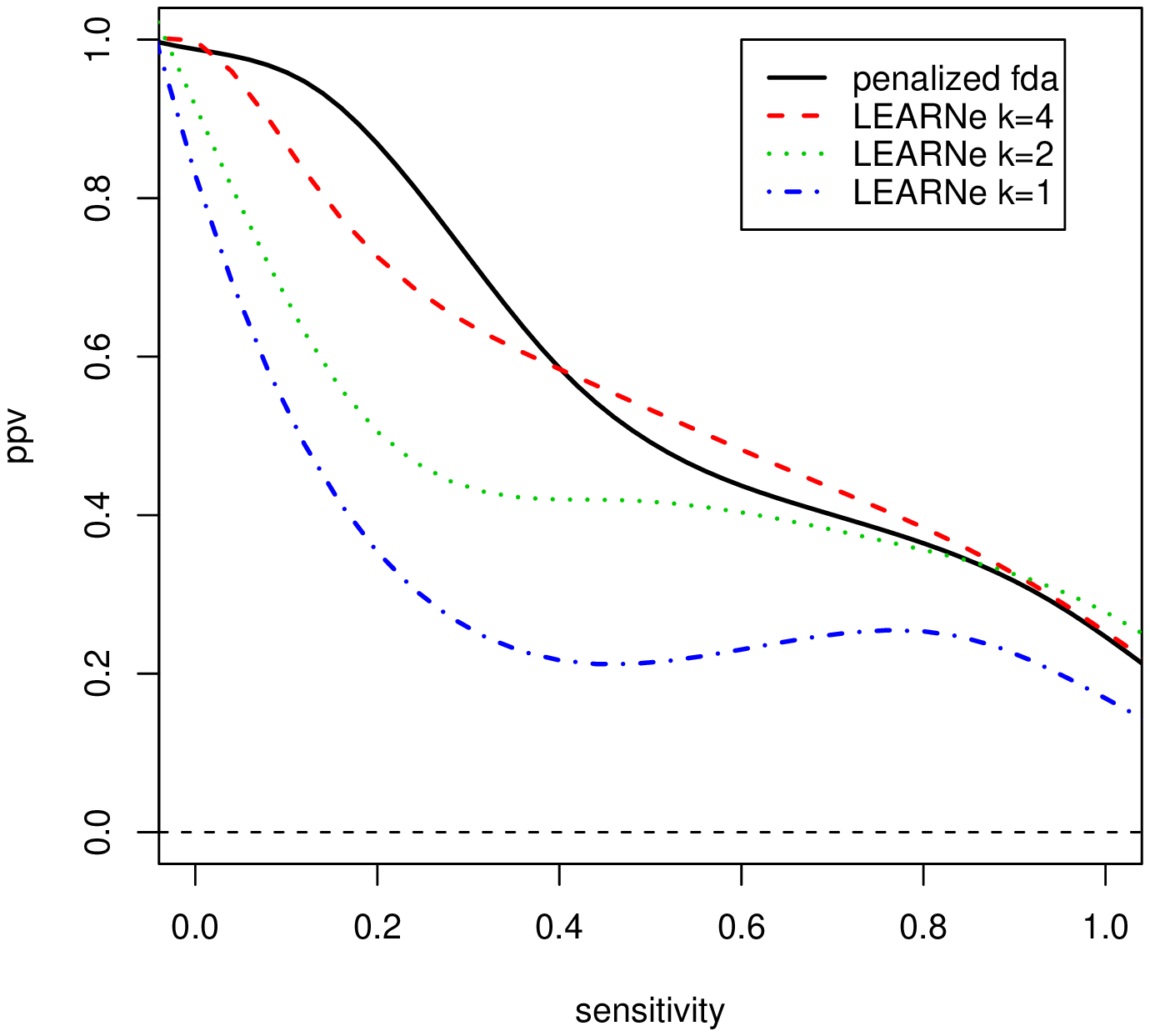}}
\caption{PPV vs. sensitivity plots for simulated networks with 10
nodes and 10 time points. The standard deviation of the noise is
$\sigma$=0.1. The algorithm LEARNe is applied with subset size
$k=4$, $k=2$ and $k=1$.}\label{fig:diffk}
\end{figure}

In the above simulations, we used $k=4$ when applying LEARNe to the
simulated data, where $k$ is the subset size to search over in
LEARNe. In these simulated data, it is known that the connection size is
actually $2$ for each gene. Curiously, as shown in Figure \ref{fig:diffk},
using $k=2$ results in much worse performances. As expected, if we
use $k=1$, the result is even worse. This shows that the
performance of LEARNe depends critically on the size of subsets searched,
which is in turn constrained by the computational resources available.

We demonstrate the performance of the our penalized functional
model with the application to the cell cycle regulatory network of
Saccharomyces cerevisiae. The dataset comes from \citep{spellman98} which
provides a comprehensive list of cell cycle regulated genes
identified by time course expression analysis. We use the 18 time
points of the alpha factor synchronized expression data. This
dataset has been used widely for evaluating a wide variety of
statistical models. Same as \cite{nam07}, we consider 20
genes including 4 transcription factors known to be involved in
regulatory functions during different stages of the cell cycle.

\begin{figure}[!tpb]
\centerline{\includegraphics[width=7cm]{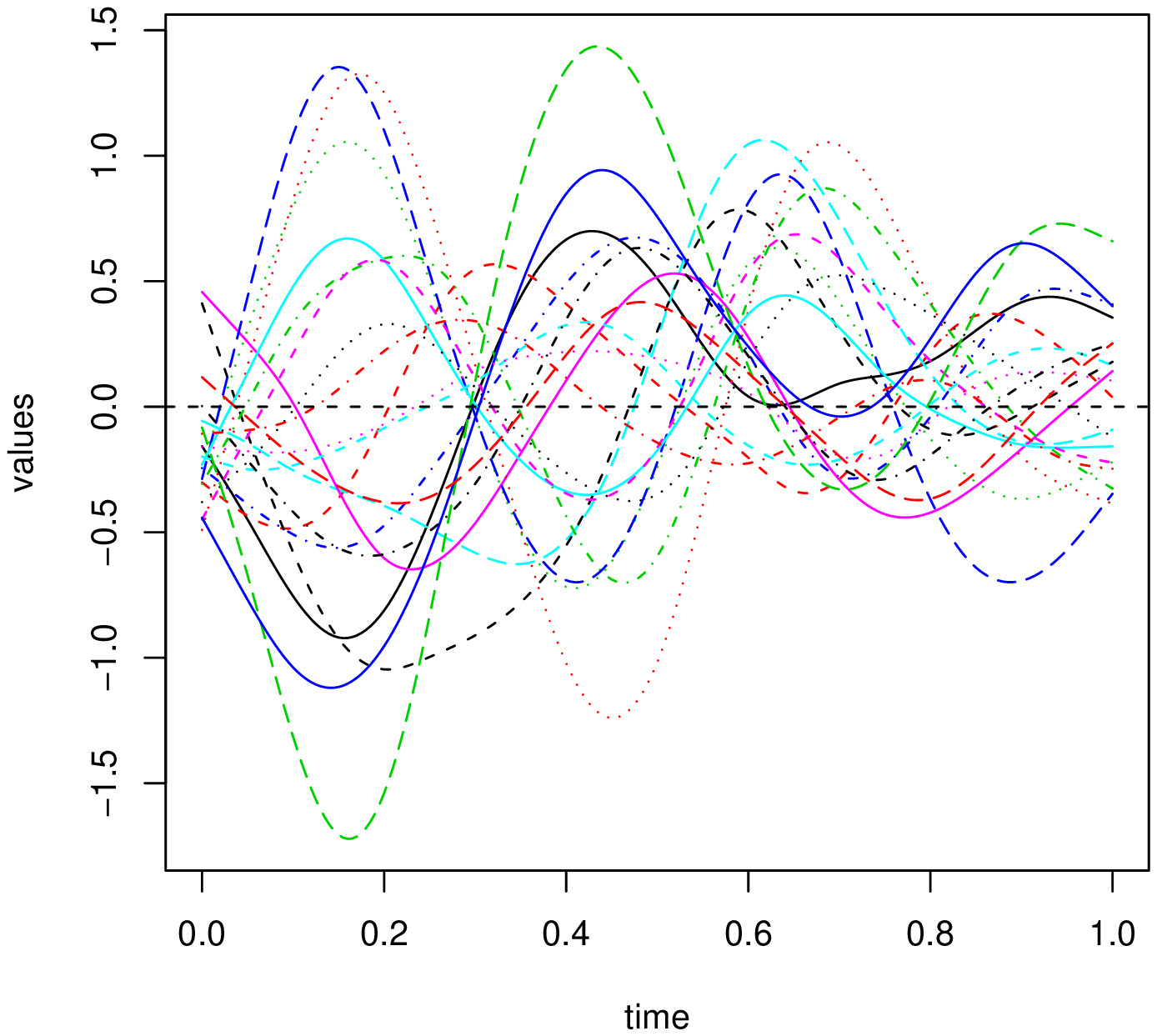}}
\caption{The expression levels of 20 genes from the cell cycle
time course expression data, represented as continuous
curves.}\label{fig:fda}
\end{figure}

We apply our approach to this dataset. The temporal evolution of
these 20 genes are shown in Figure \ref{fig:fda} after B-spline
smoothing of the expression data. To get a final model with data-based 
inference of network structure, we use the smoothing
parameter selected by cross-validation with the same smoothing
parameter for all 20 genes. The final result with the interactions
between each gene and four transcription factor is shown in Table
\ref{tab:cycle}, and we compare the result with known
interactions retrieved from the YEASTRACT database \citep{teixeira06}. For
this submatrix, we get PPV=0.54 and sensitivity=0.80. Since all
statistical models are merely mathematical approximations to the
true world, it is plausible that automatically chosen model
undersmoothes the coefficients matrix to provide a better fit to
the data. One can also manually specify the smoothing
parameter to achieved desired sparsity of the networks.

\begin{table}[!t]
\caption{The reconstructed network structure with PPV=0.54 and
Sensitivity=0.80. The interactions retrieved from database are
denoted by `$\Box$' and the interactions inferred by the model are
denoted by `$\times$'. \label{tab:cycle}}
\center{\begin{tabular}{l|cccc}\hline
      &    ace2     &    fkh1     &    swi4     &      swi5\\\hline
ace2  & $\boxtimes$ & $\boxtimes$ &             &              \\
fkh1  &             & $\boxtimes$ &             &              \\
swi4  &             & $\times$    & $\boxtimes$ &              \\
swi5  & $\times$    & $\boxtimes$ & $\times$    &   $\boxtimes$\\
sic1  & $\boxtimes$ & $\times   $ &             &   $\Box$     \\

cln3  &             &             & $\Box$      &   $\Box$     \\
far1  & $\times$    & $\times$    &             &   $\times$   \\
cln2  &             &             & $\boxtimes$ &              \\
cln1  &             & $\Box$      & $\boxtimes$ &              \\
clb6  & $\times$    &             & $\boxtimes$ &   $\times$   \\

clb5  &             &             & $\boxtimes$ &              \\
gin4  &             &             & $\boxtimes$ &              \\
swe1  &             &             & $\boxtimes$ &              \\
clb4  &             & $\boxtimes$ &             &              \\
clb2  &             & $\boxtimes$ & $\Box$      &              \\

clb1  &             & $\boxtimes$ & $\boxtimes$ &              \\
tem1  & $\times$    & $\boxtimes$ & $\times$    &              \\
apc1  &             & $\times$    &             &   $\times$   \\
spo12 &             &             &             &   $\times$   \\
cdc20 & $\times$    & $\boxtimes$ &             &   $\times$   \\
\hline
\end{tabular}}{}
\end{table}

\section{Discussion}
We described a new algorithm for network construction using time
course expression data. The algorithm is based on functional data
analysis with connection coefficients regularized by the lasso
penalty. This is a very powerful approach that makes inference for
large networks possible due to the efficient optimization
procedures previously proposed.

Our new algorithm provides several advantages over some previous
approaches. First, it achieves noise reduction by regarding the
expression data as continuous curves. High noise contained in the expression
data usually disturbs inferences when one uses finite difference to approximate
derivatives in the model. Second, the fitting procedure can be
made fully automatic with little intervention from the user. The
parameters in the model can be chosen using well studied
statistical techniques such as cross-validation. Third, the
existing optimization procedure can efficiently infer the network
structure with the whole regularization path for the coefficients simultaneously and thus
makes inference  of large networks feasible.

It is well known that biological side information can reduce the
number of false positives and false negatives. It would be
interesting to take into account such information in future work.
For example, prior knowledge on the interactions of genes can be
incorporated into the smoothing parameter so that different
interaction coefficients can be penalized differently, resulting
in adaptive lasso penalty \citep{zou06}. We expect this strategy will achieve
desired improvement on network prediction.

\section*{Funding}
This research was funded by Singapore Ministry of Education Tier 1 SUG.

\section*{Acknowledgement}
The author thanks Dr. Dougu Nam for providing the MATLAB code for
LEARNe.

\bibliographystyle{natbib}
\bibliography{papers,books}
\end{document}